\documentstyle[aps,prl]{revtex}
\begin{document}
\input epsf.sty
\twocolumn[\hsize\textwidth\columnwidth\hsize\csname %
@twocolumnfalse\endcsname
\draft
\widetext

%
%
%

\title{Average lattice symmetry and nanoscale structural correlations
in magnetoresistive manganites}

\author{V. Kiryukhin$^{1}$, T. Y. Koo$^1$, H. Ishibashi$^{1,2}$, J. P. Hill$^3$,
and S-W. Cheong$^1$}
\address{(1) Department of Physics and Astronomy, Rutgers University,
Piscataway, New Jersey 08854}
\address{(2) Department of Materials Science, Osaka Prefecture University,
Sakai, Osaka 599-8531, Japan}
\address{(3) Department of Physics, Brookhaven National Laboratory, Upton,
New York 11973}

\date{\today}
\maketitle

\begin{abstract}

We report x-ray scattering studies of nanoscale structural correlations in
the paramagnetic phases of the perovskite manganites
La$_{0.75}$(Ca$_{0.45}$Sr$_{0.55}$)$_{0.25}$MnO$_3$,
La$_{0.625}$Sr$_{0.375}$MnO$_3$, and Nd$_{0.45}$Sr$_{0.55}$MnO$_3$. We
find that these correlations are present in the orthorhombic $O$
phase in La$_{0.75}$(Ca$_{0.45}$Sr$_{0.55}$)$_{0.25}$MnO$_3$, but
they disappear abruptly at the
orthorhombic-to-rhombohedral transition in this compound. The orthorhombic
phase exhibits increased electrical resistivity and reduced ferromagnetic
coupling, in agreement with the association of the nanoscale correlations with
insulating regions. In contrast, the correlations were not detected in the
two other compounds, which exhibit rhombohedral and tetragonal phases.
Based on these results, as well as on previously published work, we
propose that the local structure of the paramagnetic phase 
correlates strongly with
the average lattice symmetry, and that the nanoscale correlations are 
an important factor distinguishing the
insulating and the metallic phases in these compounds.

\end{abstract}


\phantom{.}
]
\narrowtext

The physical mechanism underlying the magnetic-field-induced insulator-metal
transition in perovskite manganites $A_{1-x}B_x$MnO$_3$
has been the subject of intense experimental
and theoretical investigation since its rediscovery in 1993 \cite{Chahara,Rev}.
One of the motivations for such considerable attention is the
unusually large diminution of the electrical resistivity observed at the
magnetic-field-induced transition, which is now commonly referred to as
the colossal magnetoresistance (CMR). Several different kinds of the
CMR effect are known \cite{Rev}. The most widely studied variant of this 
effect is the transition from a paramagnetic insulating (PI) to a 
ferromagnetic metallic (FM) phase. The large difference between the 
resistivities of these two phases lies at the heart of the CMR effect.
The metallic nature of the FM phase has been explained within the framework of
the double-exchange mechanism \cite{Rev}.
However, the physical mechanism responsible for the
large resistivity of the PI phase 
remains poorly understood.

The situation is complicated significantly by the fact that transport
properties of the paramagnetic
state in manganites with the same doping level $x$, but
with different cations $A$ and $B$, are often very different. For example,
the electrical
resistivity of La$_{0.7}$Sr$_{0.3}$MnO$_3$ shows a metallic behavior ({\it i.e.}
grows with temperature) \cite{Uru},
while the resistivity of La$_{0.7}$Ca$_{0.3}$MnO$_3$
exhibits the temperature dependence typical of an insulator, 
decreasing with temperature \cite{lcmores}. As a consequence,
the resistivity of the latter compound is significantly larger than that
of the former. These differences cannot be explained
by steric modification of the electronic bandwidth $W$ due to the 
evolution of the average structural parameters from one composition to 
another as a result of the differing size of the divalent dopants
\cite{band1,band2}. In fact, the value of $W$ in
La$_{0.7}$Ca$_{0.3}$MnO$_3$ is expected to differ from that in 
La$_{0.7}$Sr$_{0.3}$MnO$_3$ by less than 1\% as a result of 
such effects \cite{band1}.
The widely accepted solution to this problem, first proposed theoretically
\cite{Millis}, is that the electron-phonon coupling $\lambda$ plays an
essential role. The resistivity is expected to be larger in compounds
with large values of $\lambda$. 

There is a significant amount of experimental evidence that
electron-phonon coupling plays a key role in the manganites.
It is well established that small lattice polarons are present in the PI state
\cite{P,P1,Polarons,Louca}. A lattice polaron forms when an $e_g$ electron
localizes on a Mn$^{3+}$ ion, and the surrounding oxygen octahedron distorts
due to the Jahn-Teller effect. Formation of lattice polarons leads to an 
increase of the electrical resistivity. In addition
to single polarons, the paramagnetic state exhibits
correlated lattice distortions with a correlation length of 
several lattice constants \cite{Corr}.  
It has been proposed that the structural correlations reported in 
Ref. \cite{Corr}
reflect the presence of nanoscale regions
possessing charge and orbital order, and that these regions are responsible
for the enhanced resistivity of the PI phase. The temperature dependence
of the electrical resistivity has
been shown to follow the concentration of the correlated regions \cite{Corr},
in agreement with this hypothesis. Further, the
correlations are strongly suppressed in an applied magnetic field \cite{vkirM},
implying that they play an important role in the CMR effect.
The detailed structure of the correlated
regions remains to be elucidated \cite{BJC},
but recent experimental studies suggest \cite{vkirX,Zuo} that the 
correlated regions exhibit CE-type and striped charge and orbital ordered 
structures similar to those found in manganites possessing long-range
charge and orbital order \cite{ZP}.
Numerous theoretical works have also been devoted to the problem of 
structural and electronic
inhomogeneities in manganites \cite{Dagotto}, but no realistic
description of the PI state has been achieved thus far.

In the above scenario, the differences in the transport 
properties of the PI state in various manganites must then be explained by
variations in the concentration, size, or other properties of the
correlated regions. Such variations might, in principle, result from a
dependence of the electron-phonon coupling $\lambda$ on sample composition.
However, $\lambda$ is not expected to depend significantly on the
$A$-site ionic composition (because it is, essentially, a local
parameter proportional to the Jahn-Teller energy $E_{JT}$). 
Thus, other reasons for the observed variations of physical properties must
be identified. Among the possible explanations, it has been
proposed, for example, that in manganites with a large
variance in the $A$ cation radius, cation size disorder could 
produce strain fields
stabilizing local Jahn-Teller distortions \cite{Mar}. This explanation does
not work, however, in the case of the La$_{0.7}$Sr$_{0.3}$MnO$_3$ and
La$_{0.7}$Ca$_{0.3}$MnO$_3$ compounds discussed above, since the metallic Sr
compound exhibits {\it larger A}-site cation radius variation than the Ca
compound, which is insulating at high temperatures. 

In this paper, we discuss a different scenario. Specifically, we point out that
the symmetry of the average structure can affect the energetics for the
formation of local distortions by allowing certain types of distortions to 
occur more easily, or, alternatively, by suppressing them. 
The importance of the average long-range structure stems from the fact that the
significant lattice distortions associated with the correlated regions give
rise to long-range anisotropic strain \cite{KK}. Thus,
lattice symmetry is one of the 
important factors determining how easily the local distortions are accommodated
in a given crystal structure.
Manganites exhibit structural transitions between distorted perovskite phases
possessing orthorhombic, rhombohedral, tetragonal, and monoclinic
lattice symmetry. As explained above,
it is reasonable to expect that some of these phases will
be more susceptible towards formation of local distortions than others. 

In this work, therefore,
we investigate the connection between the average lattice symmetry
and the local (nanoscale) 
correlations in several perovskite manganite compounds. 
X-ray scattering studies of the structural correlations
in manganites exhibiting orthorhombic, rhombohedral, and tetragonal structural
phases are reported. We find that in the samples investigated, as well as in
the previously studied manganites,
these correlations are present only
in the orthorhombic $O$ phase. This result is most convincingly demonstrated in
experiments with La$_{0.75}$(Ca$_{0.45}$Sr$_{0.55}$)$_{0.25}$MnO$_3$. 
This compound undergoes an 
orthorhombic-to-rhombohedral transition with increasing temperature.
We find that the structural correlations abruptly disappear at the transition
to the rhombohedral state. Based on these results and on previously
published work, we propose that the correlations form most easily in the
orthorhombic $O$ phase. In this phase, the crystal lattice is contracted
along the $c$ axis, as is also the case in structures which exhibit
charge-ordered phases,
in which the long axes of the Mn$^{3+}$O$_6$ octahedra lie in the $ab$ plane.
In contrast,
the correlations are suppressed in rhombohedral and tetragonal 
phases, in which the MnO$_6$ octahedra are either undistorted, or elongated
along the $c$ axis, on average. Thus, 
the local structure of the paramagnetic phase and,
consequently, its physical properties do, in fact, correlate strongly with
the average lattice symmetry.

\begin{figure}
\centerline{\epsfxsize=2.9in\epsfbox{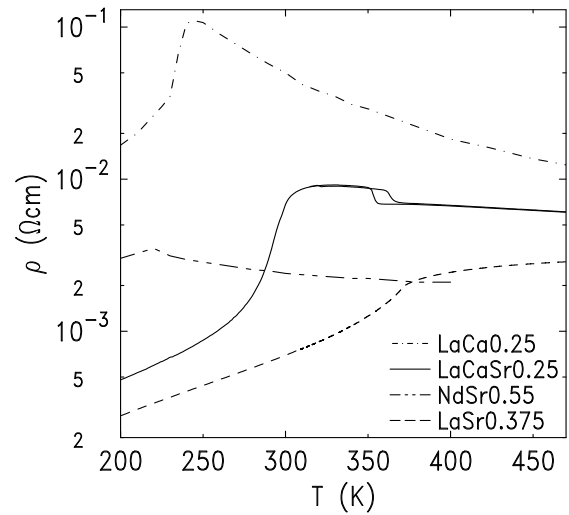}}
\vskip 5mm
\caption{Temperature dependence of the electrical resistivity of
La$_{0.75}$Ca$_{0.25}$MnO$_3$,
La$_{0.75}$(Ca$_{0.45}$Sr$_{0.55}$)$_{0.25}$MnO$_3$,
Nd$_{0.45}$Sr$_{0.55}$MnO$_3$ (taken from
Ref. [21]), and
La$_{0.625}$Sr$_{0.375}$MnO$_3$.}
\label{fig1}
\end{figure}

Single crystals of La$_{0.75}$(Ca$_{0.45}$Sr$_{0.55}$)$_{0.25}$MnO$_3$
(LCSMO), La$_{0.625}$Sr$_{0.375}$MnO$_3$, and Nd$_{0.45}$Sr$_{0.55}$MnO$_3$
were grown using the floating zone technique. X-ray diffraction
measurements were carried out at beamline X22C at the National Synchrotron
Light Source. A 10.3-keV x-ray beam was focused by a mirror, 
monochromatized by a double-crystal
Ge (111) monochromator, scattered from the sample,
and analyzed with a pyrolytic graphite crystal. The
samples were mounted in a closed-cycle refrigerator (T=6-450 K). In this 
paper, Bragg peaks are indexed in the orthorhombic $Pbnm$ notation, in which
the $a$, $b$, and $c$ axes run along the (1,1,0), (1,-1,0), and (0,0,1) cubic
perovskite directions, respectively. Scattering vectors ($h,k,l$) are given in
reciprocal lattice units. 

\begin{figure}
\centerline{\epsfxsize=2.9in\epsfbox{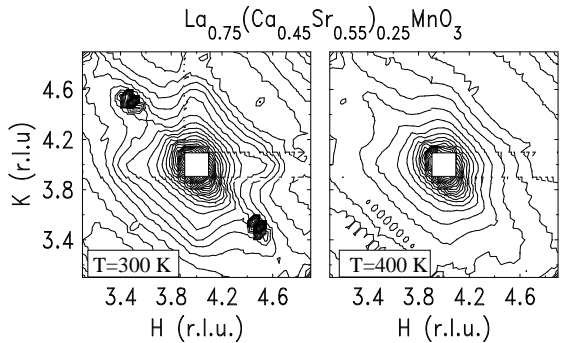}}
\vskip 5mm
\caption{Contour plots of the x-ray intensity around the (4,4,0) Bragg peak
at $T$=300 K (orthorhombic phase) and $T$=400 K (rhombohedral phase) in
La$_{0.75}$(Ca$_{0.45}$Sr$_{0.55}$)$_{0.25}$MnO$_3$.}
\label{fig2}
\end{figure}

\begin{figure}
\centerline{\epsfxsize=2.9in\epsfbox{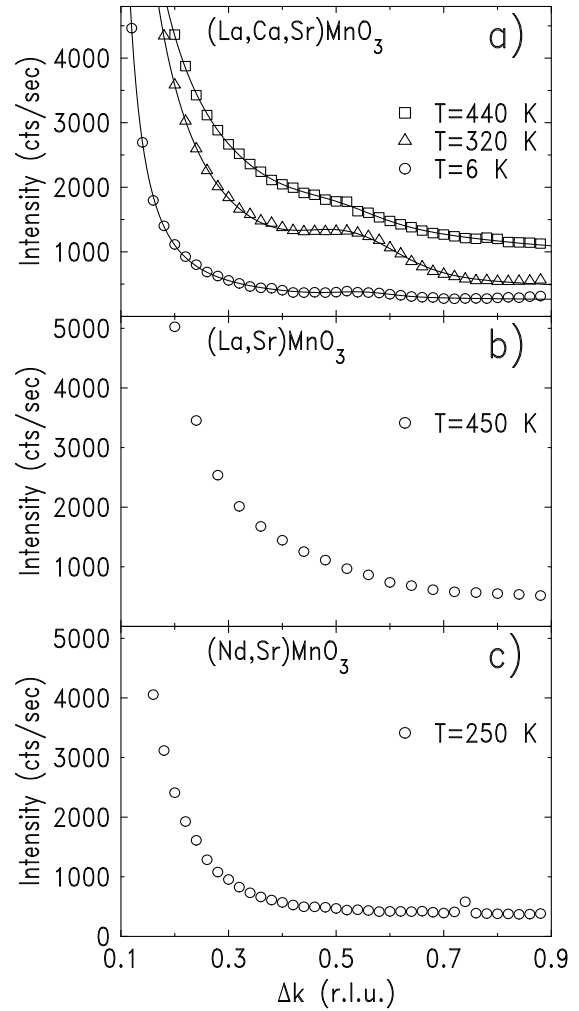}}
\vskip 5mm
\caption{X-ray scans along the (4,4+$\Delta k$,0) direction in
La$_{0.75}$(Ca$_{0.45}$Sr$_{0.55}$)$_{0.25}$MnO$_3$ (a), along
(6,2+$\Delta k$,0) in La$_{0.625}$Sr$_{0.375}$MnO$_3$ (b), and along
(6,4+$\Delta k$,0) in Nd$_{0.45}$Sr$_{0.55}$MnO$_3$ (c).
The solid lines are the results of fits, as described in the text.}
\label{fig3}
\end{figure}

We first focus our discussion on the properties of the LCSMO sample.
Its high temperature PI state exhibits rhombohedral symmetry. With decreasing
temperature, this compound first undergoes a first order
structural transition into the orthorhombic
PI phase at $T_s\approx$360 K, and then becomes a ferromagnetic metal at 
$T_c\approx$300 K. These transitions are clearly reflected in the behavior of
the electrical resistivity, shown as solid line in Fig. \ref{fig1}.  
The x-ray scattering patterns around the (4,4,0) Bragg peak in the
orthorhombic ($T$=300 K) and rhombohedral ($T$=400 K) phases are shown in 
Fig. \ref{fig2}. The sharp peaks at (4.5,3.5,0) and (3.5,4.5,0) at $T$=300 K
arise from crystallographic
twinning in the orthorhombic phase. The ``butterfly''-shaped
feature in the center is commonly attributed to scattering from uncorrelated
polarons, also known as Huang scattering, together with
thermal-diffuse scattering
\cite{Polarons}. At $T$=300 K, there are also 4 broad peaks located at
(4$\pm$0.5,4,0) and (4,4$\pm$0.5,0). These peaks are best seen in scans along
the $h$ or $k$ directions. One such scan at $T$=320 K is shown in 
Fig. \ref{fig3}(a).  

The broad peaks at (4$\pm$0.5,4,0) and (4,4$\pm$0.5,0) reflect the 
presence of nanoscale correlated regions which are believed to possess
charge and orbital order \cite{Corr}. The intensity of these peaks reflects the
concentration of the correlated regions \cite{vkirX}. To study the
temperature dependence of this concentration, we took data at different
temperatures. Some of these data are shown in Fig. \ref{fig3}(a). The data
were fitted to a sum of a Gaussian line shape and a monotonically sloping
background, the latter described by a power-law function. This background
contains contributions from single polaron and thermal-diffuse scattering
\cite{Polarons}. The error bars were estimated as described in Ref. 
\cite{vkirX}. 

\begin{figure}
\centerline{\epsfxsize=2.9in\epsfbox{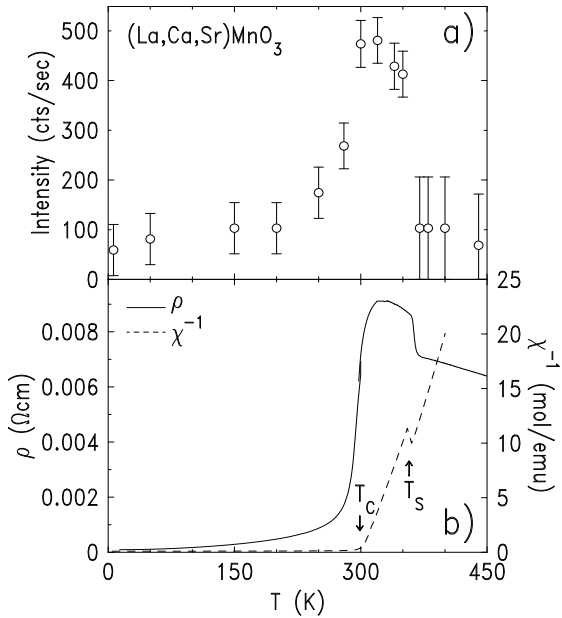}}
\vskip 5mm
\caption{(a) Temperature dependence of the intensity of the peak due to the
structural correlations in La$_{0.75}$(Ca$_{0.45}$Sr$_{0.55}$)$_{0.25}$MnO$_3$.
The single polaron background is subtracted as discussed in the text. (b)
Temperature dependence of the electrical resistivity (solid line), and inverse
magnetic susceptibility (dashed line).
The data were taken on heating.}
\label{fig4}
\end{figure}

The intensity of the correlated peak in LCSMO as a function of temperature is
shown in Fig. \ref{fig4}(a). 
The central observation
of this work is that the correlated
regions abruptly disappear at the transition from
the orthorhombic to rhombohedral phase. Thus, in LCSMO, the structural 
correlations are present only in the orthorhombic state. 
The data of Fig. \ref{fig4}(a) show that
the correlated regions also disappear in the FM state, as was reported
in earlier works \cite{Corr,vkirX}. However, in contrast to the FM state,
the rhombohedral phase is insulating. Therefore, the suppression of the
correlated insulating regions in this phase is a much more
surprising result.

Fig. \ref{fig4}(b) shows that the electrical
resistivity abruptly decreases and the magnetic susceptibility increases
as the sample enters the rhombohedral phase and the correlated regions
disappear. The Curie temperature and the effective magnetic moment extracted
from the susceptibility data are $T_c$=302(2) K, $p$=6.2, and
$T_c$=322(3) K, $p$=5.6 for the orthorhombic and the rhombohedral states,
respectively. Thus, the orthorhombic state possesses the larger effective
moment, but smaller ferromagnetic exchange constant than the corresponding
values in the rhombohedral state.
This observation is in agreement with the proposed CE-type charge and orbital
ordered structure of the correlated regions, which should be insulating and
exhibit antiferromagnetic correlations, provided that these regions are
sufficiently large \cite{vkirX,Zuo}. We note, however, that neutron scattering
experiments have thus far failed to detect the CE-type
antiferromagnetic correlations in the
PI state \cite{Corr}, and therefore the character of the magnetic interaction
in the correlated regions is yet to be established experimentally.  
Alternatively, the increased effective ferromagnetic exchange constant
of the more conducting
rhombohedral phase can stem from the effects of double exchange
which favors ferromagnetic correlations. 

The ratio of the integrated intensity of the (4,4.5,0) peak to the integrated
intensity of the main Bragg (4,4,0) peak in LCSMO is of the order of 10$^{-5}$. 
In the long-range charge-ordered phases, such as that in 
La$_{0.5}$Ca$_{0.5}$MnO$_3$, the typical intensity ratio between the 
brightest charge-ordering peaks and Bragg peaks is of the order of
10$^{-2}$ (Ref. \cite{Radaelli}).
This is significantly larger than what is observed in LCSMO.
One can estimate the intensity ratio of the (4,4.5,0) and the (4,4,0) 
peaks using the structural models proposed in Ref. \cite{Radaelli} 
(the CE-type charge and orbital ordered structure) for 
La$_{0.5}$Ca$_{0.5}$MnO$_3$, and in Ref. \cite{ZP}
(the ordering of Zener polarons) for Pr$_{0.6}$Ca$_{0.4}$MnO$_3$. 
The calculated ratios are 
2.7$\times$10$^{-2}$ for the former, and 4.1$\times$10$^{-2}$ for the latter 
model structure. Assuming that 
the correlated regions in LCSMO are small charge ordered regions with the 
structure identical to the charge-ordered state in
La$_{0.5}$Ca$_{0.5}$MnO$_3$, we obtain the result
that the volume fraction of these regions (defined here as the fraction
of MnO$_6$ octahedra experiencing the {\it correlated} distortion)
is smaller than 0.1\%. In contrast, a much larger estimate
of ``a few percent'' was obtained for the same fraction in a recent
electron diffraction study of La$_{2/3}$Ca$_{1/3}$MnO$_3$ \cite{Zuo}.
The reason for such a discrepancy is currently unclear. In either case,
it is hard to see how such an apparently
small volume fraction of the charge and orbital ordered insulating phase
could significantly affect the transport properties. 

Thus, these observations appear to contradict the strong correlation between
the nanoscale correlations and the
transport properties observed 
in a large number of manganites of different chemical
compositions and doping levels \cite{Corr,vkirM,vkirX,Zuo}.
One way to explain this contradiction is to assume
that the lattice distortions are significantly (at least a factor of 10)
smaller in the correlated regions than they are
in the corresponding long-range ordered phases.
Since the diffraction signal from the correlated
regions is proportional
to the square of the magnitude of the lattice distortion, much larger
volume fractions of the charge and orbital ordered
insulating phase are then deduced from 
the experimental data under this assumption. 

Alternatively, more elaborate models 
of the inhomogeneous PI state should be proposed. In particular, dynamic
inhomogeneities might play an important role in this state, as evidenced by 
recent experiments in the related bi-layered manganites \cite{Glass},
in which the structural correlations were found to become completely dynamic
at high temperatures. Also, the role of the {\it uncorrelated} local
distortions must be clarified. The latter distortions are known to be 
present in both the rhombohedral and the orthorhombic manganites, and to have
composition-independent distortion magnitude \cite{Louca}. In the orthorhombic 
La$_{1-x}$Ca$_x$MnO$_3$ system,
the number of MnO$_6$ octahedra experiencing a distortion (correlated or
otherwise) was found to be functionally
correlated to the magnetic and transport properties \cite{P1,P2}. The relative
contributions of the correlated and uncorrelated distortions to these functional
relationships remain to be characterized. It also remains to be seen whether
these relationships hold on the rhombohedral side of the structural
transition boundary. In this work, we do not attempt to characterize the
uncorrelated distortions because it is difficult to distinguish 
scattering due to these distortions from
thermal-diffuse scattering in our energy-integrating measurements.
Thus, while it is clear that the correlations play an
important role in the manganites, more work is needed to elucidate the
microscopic structure of the PI state.

We next briefly discuss the other samples. La$_{0.625}$Sr$_{0.375}$MnO$_3$
is rhombohedral (space group $R\bar 3c$)
at all temperatures and exhibits a paramagnetic state for
$T>$370 K. The paramagnetic insulating state of 
Nd$_{0.45}$Sr$_{0.55}$MnO$_3$ is tetragonal (space group
$I4/mcm$) \cite{Yoshi}. In the latter sample, the PI state is observed
for $T>T_N\approx$220 K. The electrical resistivities of these samples are
shown in Fig. \ref{fig1}. In the both samples, we did not detect any
evidence of the structural correlations at any temperature. 
Figs. \ref{fig3}(b,c)
show some of the collected data for these samples (Bragg peaks are again indexed
using the orthorhombic notation). In contrast, these correlations are
clearly present in the paramagnetic orthorhombic phase of
La$_{0.75}$(Ca$_{0.45}$Sr$_{0.55}$)$_{0.25}$MnO$_3$ discussed above, 
and also in the PI
phases of numerous other
orthorhombic manganites, such as
Nd$_{1-x}$Sr$_x$MnO$_3$
(0.3$\leq x\leq$0.5) \cite{vkirX}, La$_{1-x}$Ca$_x$MnO$_3$ 
(0.2$\leq x\leq$0.3) \cite{Corr}, and Pr$_{1-x}$Ca$_x$MnO$_3$ 
(0.3$\leq x\leq$0.5) \cite{Pr}.
We suggest then that the nanoscale structural
correlations described above can {\it only} occur in an orthorhombic phase. 
Of course, more systematic studies are needed to verify this intriguing
hypothesis, but even absent such studies, we feel that there is sufficient
evidence in hand to warrant speculation as to why such an observation
might be true.

All the orthorhombic samples discussed above exhibit the well known 
$O$ structure with the lattice constants $c/\sqrt 2<b\sim a$ 
\cite{Yoshi,QHuang}.
An orthorhombic state with $c/\sqrt 2<b\sim a$
is also realized in the manganites with long-range CE-type charge and orbital
order. In this structure, the $c$-axis contraction reflects the global effect
of ordered
Jahn-Teller distortions of the Mn$^{3+}$O$_6$ octahedra with the long axis
lying in the $ab$ plane \cite{Rev}. Thus, on average, the long axes of the
MnO$_6$ octahedra lie in the $ab$ plane 
in both the long-range charge ordered phase,
and in the $O$ paramagnetic insulating state \cite{Yoshi,QHuang}. 
In contrast, the MnO$_6$ octahedra are undistorted in the rhombohedral
$R\bar 3c$ state \cite{Mitchell}, and are significantly elongated along the
$c$-axis in the tetragonal $I4/mcm$ state \cite{Yoshi}. 

It is not completely
surprising that the MnO$_6$ octahedra are distorted in a similar manner
in the manganites exhibiting nanoscale charge-ordered
regions and in the manganites with long-range charge and orbital order. 
Note, however, that this is a quite nontrivial result because
there is no {\it a priori} requirement that the local symmetry of the nanoscale
ordered regions should match the average symmetry of the crystal lattice.
The energetically favorable
response of the orthorhombic structure to the long-range strains 
produced by the local distortions is likely to be one of the key factors
explaining the reported observations. Theoretical calculations taking into
account the details of the lattice structure are needed to confirm these ideas.
In any case, the proposed correlation between
the presence of the local lattice distortions and the average crystallographic
structure is intriguing and deserves further theoretical and
experimental investigation.

Finally, we return to the problem of the drastic variation of the electrical
properties of the paramagnetic phase with chemical composition. 
We emphasize that here we are discussing the
high-temperature paramagnetic phase, and not the low-temperature state.
Specifically,
we will consider the La$_{1-x}$(Ca$_{1-y}$Sr$_{y}$)$_{x}$MnO$_3$ series of
compounds with $x$ in the vicinity of 0.3. This system exhibits a well
defined insulating behavior for $y$=0 (see Fig. \ref{fig1}), and is metallic
for $y$=1 (Fig. \ref{fig1}, and Ref. \cite{Uru}). 
At fixed $x$ and with $y$
increasing from 0 to 1, the system should, therefore, undergo an 
insulator-to-metal transition.
It is well known that with $y$ increasing from 0 to 1, this system also
undergoes an
orthorhombic-to-rhombohedral structural transition \cite{PhaseD}.  
With the exception of the narrow region in the vicinity of the structural
transition, the electrical resistivity shows the temperature dependence
typical for an insulator in the orthorhombic samples, and that of a metal
in the rhombohedral state \cite{PhaseD}. These data show that
the metal-insulator transition coincides approximately with the structural
transition in this series of compounds. 

In the vicinity of the structural transition, the electrical 
resistivity exhibits temperature-dependent behavior typical of an
insulator in the both structural phases, see Fig. \ref{fig4}(b), and Ref.
\cite{PhaseD}. We would like to point out an interesting possibility that 
the insulating behavior of the rhombohedral phase 
could result from the effects of chemical and magnetic 
disorder. These effects are, in fact, well known to play an important role in
mixed-valence manganites \cite{Rev}.
In the La$_{1-x}$(Ca$_{1-y}$Sr$_{y}$)$_{x}$MnO$_3$ series of compounds,
in particular, the importance of disorder is
evidenced by the large magnitude of the electrical resistivity
deep in the paramagnetic metallic region of the phase diagram \cite{PhaseD}.
It is, therefore, reasonable to expect that in samples with less chemical
disorder, the metal-insulator transition boundary lies closer or even
coincides with the structural
transition boundary. We, therefore, believe that further theoretical and
experimental investigation of the role of disorder in the 
La$_{1-x}$(Ca$_{1-y}$Sr$_{y}$)$_{x}$MnO$_3$ series of compounds is of great
interest. 

The experimental data for one of the compounds from this series,
La$_{0.75}$(Ca$_{0.45}$Sr$_{0.55}$)$_{0.25}$MnO$_3$, were discussed in detail
above. Our experiments clearly show that the nanoscale 
structural correlations are 
present only on the orthorhombic side of the phase boundary. In addition,
these correlations are clearly present in the $y$=0
compound La$_{0.75}$Ca$_{0.25}$MnO$_3$ \cite{Corr}, and are undetectable in the
$y$=1 material La$_{0.625}$Sr$_{0.375}$MnO$_3$. Thus, the existing
experimental data show
that the metallic rhombohedral state in this compound series does not
exhibit the nanoscale correlations, while the correlations are present in
the insulating orthorhombic state. The presence of the nanoscale insulating
regions, therefore, appears to be an important factor distinguishing the
insulating and the metallic phases. Thus, the nanoscale correlations should 
be included in any physical mechanism explaining the observed large variation
of the high-temperature properties of manganites and, consequently,
the observed values of the CMR effect.

In summary, we report x-ray scattering studies of nanoscale structural
correlations in several manganite compounds exhibiting orthorhombic,
rhombohedral, and tetragonal perovskite 
phases. Based on these results, and also 
on previously published work, we propose that these correlations form most
easily in the orthorhombic $O$ phase, and are suppressed in the rhombohedral
and the tetragonal phases. Thus, the local structure of the paramagnetic
phase appears to correlate strongly with the average lattice symmetry. 
Our experiments strongly suggest
that changes in physical properties 
observed at structural transitions in manganites
cannot be understood through spatially
uniform phases, and that nanoscale inhomogeneities must play an essential role
in these transitions.

{\it Note added:} A recent work by Mira {\it at al.} reports measurements
of thermal properties, magnetoresistance, and lattice expansion in
various manganites \cite{Mira}. 
The analysis of these quantities led the authors of Ref. \cite{Mira} to a
conclusion that lattice effects (electron-lattice interaction, and local 
Jahn-Teller distortions) are important in the orthorhombic manganites,
while these effects are strongly reduced in the rhombohedral samples.  
The orthorhombic
PI state was proposed to contain a superparamagnetic secondary phase in
a certain temperature interval above $T_c$. These results are consistent
with the x-ray diffraction measurements reported in our paper, providing
further evidence of the correlation between the local
structure and the average lattice symmetry, and thus revealing the 
physics underlying the drastic differences in the properties of manganites
with different lattice symmetry.

We are grateful to K. I. Kugel,
A. J. Millis, and J. Mira for important discussions. This work was
supported by the NSF under grants No. DMR-0093143, and DMR-0103858, by the 
DOE under Contract No. DE-AC02-98CH10886, 
and by the NSF MRSEC program, Grant No.
DMR-0080008.



\begin{references}

\bibitem{Chahara} S. Chahara, T. Ohno, K. Kasai, Y. Kozono, Appl. Phys. Lett. 
{\bf 63}, 1990 (1993); R. von Helmolt, J. Wecker, B. Holzapfel, L. Schultz,
and K. Samwer, Phys. Rev. Lett. {\bf 71}, 2331 (1993);
S. Jin, T. H. Tiefel, M. McCormack, R. A. Fastnacht,
R. Ramesh, and L. H. Chen, Science {\bf 264}, 413 (1994).

\bibitem{Rev} For a review, see {\it Colossal Magnetoresistance Oxides}, edited
by Y. Tokura (Gordon and Breach, London, 1999).

\bibitem{Uru} A. Urushibara, Y. Moritomo, T. Arima, A. Asamitsu, G. Kido,
and Y. Tokura, Phys. Rev. B {\bf 51}, 14103 (1995).

\bibitem{lcmores} P. Dai, H. Y. Hwang, J. Zhang, J. A. Fernandez-Baca,
S-W. Cheong, C. Kloc, Y. Tomioka, and Y. Tokura, Phys. Rev. B {\bf 61},
9553 (2000).

\bibitem{band1} P. G. Radaelli, G. Iannone, M. Marezio, H. Y. Hwang, 
S-W. Cheong, J. D. Jorgensen, and D. N. Argyriou, Phys. Rev. B
{\bf 56}, 8265 (1997).

\bibitem{band2} V. Laukhin, J. Fontcuberta, J. L. Garcia-Mu\~noz, and
X. Obradors, Phys. Rev. B {\bf 56}, R10009 (1997).

\bibitem{Millis} A. J. Millis, P. B. Littlewood, and B. I. Shraiman, Phys. Rev.
Lett. 74, 5144 (1995); A. J. Millis, B. I. Shraiman, and R. Mueller,
{\it ibid.} {\bf 77}, 175 (1996).

\bibitem{P} S. J. L. Billinge, R. G. DiFrancesco, G. H. Kwei,
J. J. Neumeier, and J. D. Thompson, Phys. Rev. Lett.
{\bf 77}, 715 (1996);
K. H. Kim, J. Y. Gu, H. S. Choi, G. W. Park, and T. W. Noh,
{\it ibid.} {\bf 77}, 1877 (1996); M. Jaime, H. T. Hardner, M. B. Salamon,
M. Rubinstein, P. Dorsey, and D. Emin,
{\it ibid.} {\bf 78}, 951 (1997).

\bibitem{P1} C. H. Booth, F. Bridges, G. H. Kwei, J. M. Lawrence,
A. L. Cornelius, and J. J. Neumeier, Phys. Rev. Lett. {\bf 80}, 853 (1998).

\bibitem{Polarons} S. Shimomura, N. Wakabayashi, H. Kuwahara,
and Y. Tokura, Phys. Rev. Lett.
{\bf 83}, 4389 (1999); L. Vasiliu-Doloc, S. Rosenkranz, R. Osborn,
S. K. Sinha, J. W. Lynn, J. Mesot, O. H. Seeck, G. Preosti,
A. J. Fedro, and J. F. Mitchell, {\it ibid.}
{\bf 83}, 4393 (1999).

\bibitem{Louca} D. Louca, T. Egami, E. L. Brosha, H. R\"oder, and A. R. Bishop,
Phys. Rev. B {\bf 56}, R8475 (1997).

\bibitem{Corr} P. Dai, J. A. Fernandez-Baca, N. Wakabayashi, E. W. Plummer,
Y. Tomioka, and Y. Tokura, Phys. Rev. Lett. {\bf 85}, 2553 (2000);
C. P. Adams, J. W. Lynn, Y. M. Mukovskii, A. A. Arsenov, and
D. A. Shulyatev, {\it ibid.} {\bf 85}, 3954 (2000);
C. S. Nelson, M. v. Zimmermann, Y. J. Kim, J. P. Hill,
Doon Gibbs, V. Kiryukhin, T. Y. Koo, S-W. Cheong, D. Casa, B. Keimer,
Y. Tomioka, Y. Tokura, T. Gog, and C. T. Venkataraman, Phys. Rev. B
{\bf 64}, 174405 (2001).

\bibitem{vkirM} T. Y. Koo, V. Kiryukhin, P. A. Sharma, J. P. Hill, and
S-W. Cheong, Phys. Rev. B {\bf 64}, 220405(R) (2001).

\bibitem{BJC} Nanoscale structural correlations
have also been found recently in bi-layered manganites, and they were associated
with charge-density-wave fluctuations. 
However, the wave vector of these correlations is very different (rotated
45$^\circ$) from the
wave vector of the correlations in the perovskite manganites discussed here, and
therefore it is unclear whether these two phenomena are related.
See B. J. Campbell, R. Osborn, D. N. 
Argyriou, L. Vasiliu-Doloc, J. F. Mitchell, S. K. Sinha, U. Ruett, C. D. Ling,
Z. Islam, and J. W. Lynn, Phys. Rev. B {\bf 65}, 014427 (2001).

\bibitem{vkirX} V. Kiryukhin, T. Y. Koo, A. Borissov, Y. J. Kim, C. S. Nelson,
J. P. Hill, D. Gibbs, and S-W. Cheong, Phys. Rev. B {\bf 65}, 094421 (2002).

\bibitem{Zuo} J. M. Zuo, and J. Tao, Phys. Rev. B {\bf 63}, 060407(R), (2001).

\bibitem{ZP} Note, an alternative model describing the low-temperature state of 
half-doped manganites as ordering of Zener polarons has been proposed recently,
see A. Daoud-Aladine, J. Rodriguez-Carvahal, L. Pinsard-Gaudart, M. T.
Fern\'andez-Diaz, and A. Revcolevschi, Phys. Rev. Lett. {\bf 89},
097205 (2002). We, however, keep referring to this state as CE-type charge and
orbital ordered, because this assignment is dominant in the current
literature.

\bibitem{Dagotto} For a review, see E. Dagotto, T. Hotta, and A. Moreo,
Phys. Rep. {\bf 344}, 1 (2001).

\bibitem{Mar} L. M. Rodriguez-Martinez, and J. P. Attfield, Phys. Rev. B
{\bf 54}, 15622 (1996).

\bibitem{KK} D. I. Khomskii, and K. I. Kugel, Europhys. Lett. {\bf 55},
208 (2001).

\bibitem{MRS} H. Kuwahara, T. Okuda, Y. Tomioka, A. Asamitsu, and Y. Tokura,
in {\it Science and Technology of Magnetic Oxides}, edited by M. Hundley,
J. Nickel, R. Ramesh, and Y. Tokura, MRS Symp. Proc. No. 494 (Materials
Research Society, Pittsburgh, 1998), p. 83.

\bibitem{Radaelli} P. G. Radaelli, D. E. Cox, M. Marezio, and S-W. Cheong,
Phys. Rev. B {\bf 55}, 3015 (1997).

\bibitem{Glass} D. N. Argyriou, J. W. Lynn, R. Osborn, B. Campbell,
J. F. Mitchell, U. Ruett, H. N. Bordallo, A. Wildes, and C. D. Ling,
Phys. Rev. Lett. {\bf 89}, 036401 (2002).

\bibitem{P2} M. Jaime, P. Lin, S. H. Chun, and M. B. Salamon, Phys. Rev. B
{\bf 60}, 1028 (1999).

\bibitem{Pr}
S. Shimomura, T. Tonegawa, K. Tajima, N. Wakabayashi, N. Ikeda,
T. Shobu, Y. Noda, Y. Tomioka, and Y. Tokura, Phys. Rev. B {\bf 62}, 3875
(2000); R. Kajimoto, H. Yoshizawa, Y. Tomioka, and Y. Tokura, {\it ibid.}
{\bf 63}, 212407 (2001).

\bibitem{Yoshi} R. Kajimoto, H. Yoshizawa, H. Kawano, H. Kuwahara, Y. Tokura,
K. Ohoyama, and M. Ohashi, Phys. Rev. B {\bf 60}, 9506 (1999).

\bibitem{QHuang} Q. Huang, A. Santoro, J. W. Lynn, R. W. Erwin, J. A. Borchers,
J. L. Peng, K. Ghosh, and R. L. Greene, Phys. Rev. B {\bf 58}, 2684 (1998).

\bibitem{Mitchell} J. F. Mitchell, D. N. Argyriou, C. D. Potter, D. G. Hinks,
J. D. Jorgensen, and S. D. Bader, Phys. Rev. B {\bf 54}, 6172 (1996).

\bibitem{PhaseD} Y. Tomioka, A. Asamitsu, and Y. Tokura, Phys. Rev. B {\bf 63},
024421 (2000).

\bibitem{Mira} J. Mira, J. Rivas, L. E. Hueso, F. Rivadulla, M. A. L\'opez
Quintela, N. A. Se\~naris Rodriguez, C. A. Ramos, Phys. Rev. B {\bf 65},
024418 (2001).

\end{references}
\end{document}